\documentclass[review]{elsarticle}

\usepackage{lineno,hyperref}
\modulolinenumbers[5]

\journal{Journal of \LaTeX\ Templates}

\usepackage{caption}
\usepackage{natbib} 
\usepackage[T1]{fontenc}    
\usepackage{url}            
\usepackage{booktabs}       
\usepackage{amsfonts}       
\usepackage{nicefrac}       
\usepackage{microtype}      
\usepackage{amsfonts,amsmath,amssymb,amsthm,bm}
\usepackage{algorithm}
\usepackage{algorithmic}
\usepackage{graphicx}
\usepackage{color}

\newcommand{\R}{\mathbb{R}}                                     
\newcommand{\C}{\mathbb{C}}                                     

\newcommand{\vectorize}{{\rm vec}}
\newcommand{\bi}{\begin{itemize}}
\newcommand{\ei}{\end{itemize}}
\newcommand{\ba}{\begin{array}}
\newcommand{\ea}{\end{array}}

\begin{document}

\begin{frontmatter}

\title{Physically-interpretable classification of network dynamics for complex collective motions}

\author[1,2]{Keisuke Fujii\corref{mycorrespondingauthor}} 
\cortext[mycorrespondingauthor]{Corresponding author}
\ead{fujii@i.nagoya-u.ac.jp}
\author[2]{Naoya Takeishi}
\author[2]{Motokazu Hojo}
\author[3]{Yuki Inaba}
\author[4,2]{Yoshinobu Kawahara}
 
\address[1]{Graduate School of Informatics, Nagoya University}
\address[2]{RIKEN Center for Advanced Intelligence Project}
\address[3]{Japan Institute of Sports Sciences}
\address[4]{Institute of Mathematics for Industry, Kyushu University}

\begin{abstract} 
Understanding complex network dynamics is a fundamental issue in various scientific and engineering fields. 
Network theory is capable of revealing the relationship between elements and their propagation; however, for complex collective motions, the network properties often transiently and complexly change. 
A fundamental question addressed here pertains to the classification of collective motion network based on physically-interpretable dynamical properties. 
Here we apply a data-driven spectral analysis called graph dynamic mode decomposition, which obtains the dynamical properties for collective motion classification.
Using a ballgame as an example, we classified the strategic collective motions in different global behaviours and discovered that, in addition to the physical properties, the contextual node information was critical for classification.
Furthermore, we discovered the label-specific stronger spectra in the relationship among the nearest agents, providing physical and semantic interpretations. 
Our approach contributes to the understanding of complex networks involving collective motions from the perspective of nonlinear dynamical systems.
\end{abstract}

\end{frontmatter}

\section*{Introduction}
Complex systems are modelled as a collection of the discrete elements that non-linearly interact. 
Dynamic processes for such complex systems are of great interest in a variety of scientific fields, such as 
sociology\cite{Vespignani09,Holme04}, 
epidemiology\cite{Read08,Bansal10}, 
neuroscience\cite{Bullmore09,Breakspear17} 
and physics\cite{Colizza07,Holme12}. 
These systems are often represented as networked systems using graphs, which have several problems such as in
classification\cite{Newman04,Grady12}, 
prediction\cite{Choi17,Rosenthal15} 
and control\cite{Wang10,Zhou17}. 
Among various networked systems, dynamic networks of collective motions\cite{Croft04,Tanner04,Bode11},
in which nodes and links are agents and their relationships (e.g. based on their positions),
pose several challenges.
These challenges arise in changes of the relation among agents (i.e. the structure of a graph) in transient and complex ways, with the exception of moderate changes such as diffusion or contagion\cite{Rosenthal15}.
In the network dynamics of collective motions, if the governing equation is given such as mathematical models (e.g. cellular automata\cite{Mortveit07} and coupled dynamical systems\cite{Wu05}) and controlled systems\cite{Cliff16}, 
these are often modeled as graph dynamical systems (GDSs).
However, for many biological collective motions\cite{Hutchins90,Fujii16}, the relation among agents transiently and complexly changes according to different situations, with the exception of the simply modelled collective motions\cite{Aoki82,Helbing95}.
These biological network dynamics can be regarded as the network dynamics or GDSs of collective motions represented by graph sequence data.
To address this problem, conventional approaches have basically computed the properties of a graph in each temporal snapshot\cite{Centola07,Bullmore09} or in a temporal sliding window\cite{Ide04} of the sequence data.
However, for the understanding of network dynamics, these approaches are difficult to directly extract the dynamical properties, i.e. physically-interpretable information about the dynamics such as frequencies with decay/growth rate and the corresponding spatial (network) structures. 
The fundamental question addressed here is how complex collective motion networks should be classified based on the physically-interpretable dynamical properties.

The motivation of this paper is to understand network dynamics (or underlying global dynamics of GDSs) of collective motions by directly extracting the dynamical properties of the network in a data-driven manner.
As a method of describing nonlinear dynamical systems with a global mode by the direct extraction of dynamical properties, operator-theoretic approaches have attracted attention in fields such as applied mathematics, physics and machine learning. 
One of the approaches is based on the composition operator (often referred to as the Koopman operator\cite{Koopman31,Mezic05}), which defines the time evolution of observation functions in a function space, rather than directly defines the time evolution in a state space from a classical and popular view of the analysis.
The advantage of using the operator-theoretic approach is to lift the analysis of nonlinear dynamical systems to a linear (but infinite-dimensional) regime, which is more amenable to  subsequent analysis.

Among several estimation methods, one of the most popular algorithms for spectral analysis of the Koopman operator is dynamic mode decomposition (DMD), which was originally developed in fluid physics\cite{Rowley09,Schmid10}.
DMD has been successfully applied in many real-world problems, such as analyses of epidemiology and electrocorticography\cite{Proctor15,Brunton16a}. 
Among several variants of DMDs (for details, see Material and Methods), {\it Graph DMD}\cite{Fujii19b} can extract and visualise the underlying low-dimensional global dynamics of GDSs with structures among observables from data. 
However, for more general complex collective motions including strategic human groups, agents flexibly change the rules of behaviour according to the situation; thus, the computation of a feature for the classifier using the obtained spectra cannot be uniquely determined. 
Therefore, in this paper, we computed Graph DMD for each temporal sliding window and the feature for the classifier reflecting the essential (e.g. dynamical) information for multiple types of collective motions. 
Previous other approaches have extracted linear dynamics in complex networks\cite{Delvenne15} and physical interaction networks\cite{Nitzan17}. 
Recent neural network approaches automatically extract graph (spatial) and temporal information and perform classification or prediction\cite{Yan18,Yu18}.
In contrast to these approaches, our methodological contribution is to provide a classification method by directly extracting physically-interpretable dynamical properties for nonlinear GDSs in a data-driven manner.
Additionally, we propose a more straightforward formulation for Graph DMD in an observed data space than the previous one\cite{Fujii19b}.

One of the goals of this data-driven or equation-free method\cite{Fujii18} is characterising complex collective motions that emerge from complex rules, rather than motions that emerge from fewer rules such as biological rhythms (e.g. a day and year) and simple rules (e.g. attraction and repulsion).
Organised human tasks in small groups such as navigation\cite{Hutchins90} and ballgame teams\cite{Fujii16} provide excellent examples of complex dynamics and pose challenges in many research fields because of their switching and overlapping hierarchical subsystems\cite{Fujii16}. 
In these cases, the focusing networks are often small (e.g. less than 30 nodes) but spatio-temporally complex due to the highly-strategic (or contextual) behaviours.
Thus, we hypothesised that the node information (i.e. individual agent) is more critical for the understanding the subtle yet significant differences between motions \cite{Fujii16} rather than the network properties such as graph spectrum\cite{Chung97}. 
Previous works\cite{Fujii17,Fujii18} have predictively classified the time-varying interactions into two group outcomes (i.e. scored or unscored) while reflecting the time-varying interactions among attackers, defenders and the ball.
However, since the previous method utilising so-called kernel methods decompose the modes in infinite functional space to acquire high expressiveness\cite{Kawahara16}, it is difficult to physically interpret or visualise the decomposed modes in the observed data space.
Thus, the scientific contribution of this study is to provide a method for interpreting the discovered network structures corresponding to the extracted dynamical properties.
Additionally, from the viewpoint of practical applications,  supervisors such as coaches and researchers who analyse collective motions of agents (e.g. humans, other animals and objects), must exert great effort to observe the motions and  label them to specific formations or objectives.
The practical advantage of our approach is thus to automatically classify such complex collective motions to aid further analyses such as discovering network structures or dynamical properties.

As already mentioned, the purpose of this study is to understand the network dynamics for complex collective motions via physically-interpretable classification using data-driven spectral analysis of GDSs known as Graph DMD.  
To this end, we first present the theoretical framework and results of applying graph DMD to actual ballgame data.
Then, to validate our approach, we present the classification results of multiple recognition tasks for globally different collective behaviours.
Again, we hypothesise that, in addition to physical information, the contextual node information is critical for the classification.
We investigated this hypothesis by comparing various existing classification methods.
Finally, we visualise and physically and semantically interpret the network structures (called Graph DMD modes) that correspond to extracted dynamics properties.

\section*{Results}
\subsection*{Graph DMD framework.}

\begin{figure}[!t]
\centerline{\includegraphics[width=1 \textwidth]{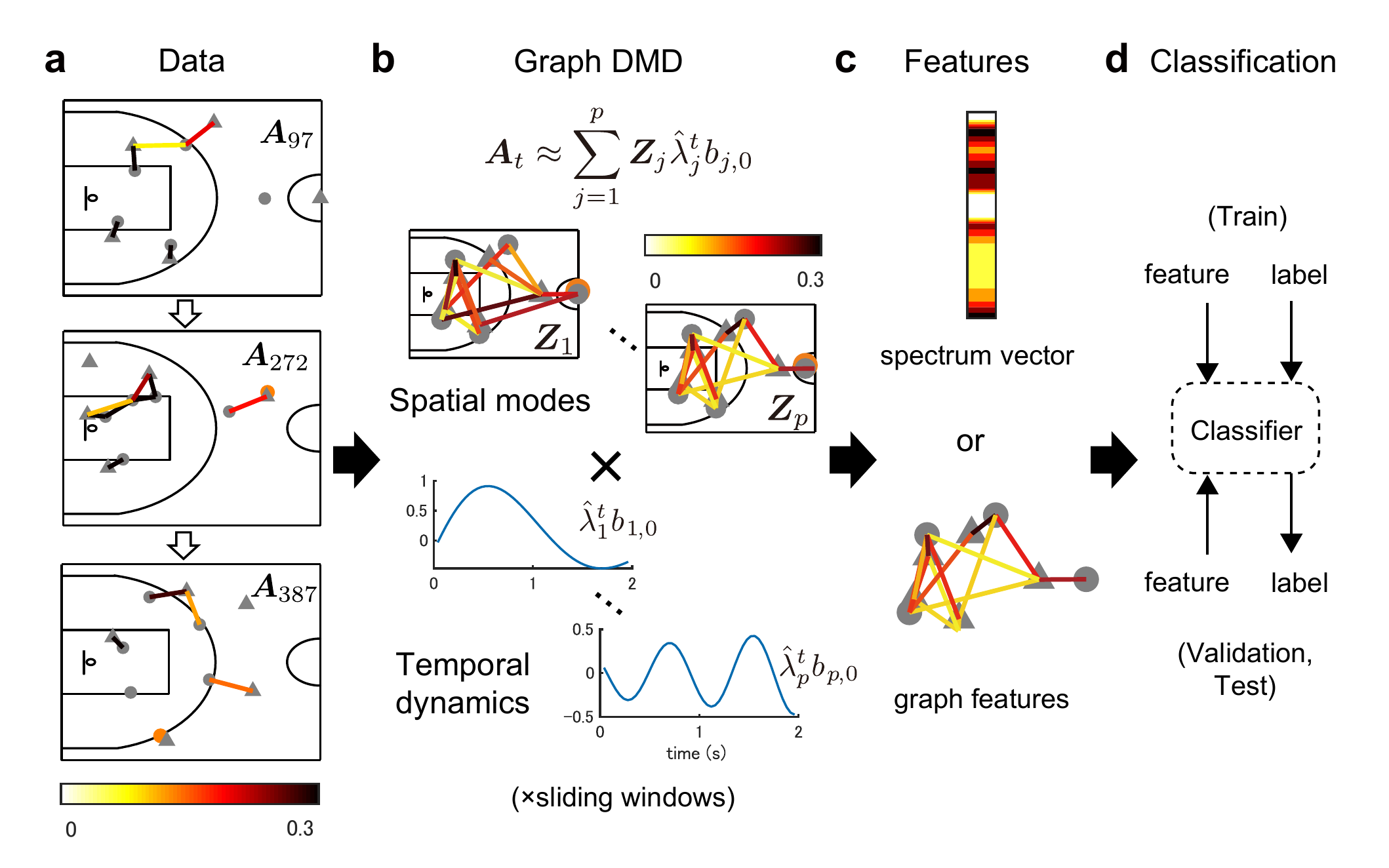}} 
\caption{{\bf Schematic diagram of Graph DMD and classification.}
(a) Obtained data in three temporal frames. The colours in each edge represent the values as a function of the distance between agents; they approach 0 if far and 1 if near (further details are provided in the main text). The input of Graph DMD is adjacency matrix series $\bm{A}_t$. (b) Graph DMD decomposes the input $\bm{A}_t$ into the sum of the product of graph (spatial) modes and temporal dynamics. To analyse the time-varying dynamics,  Graph DMD is performed for each temporal sliding window. (c) After Graph DMD, features for classification are computed in different approaches. One is simply to vectorise the Graph DMD modes (i.e. spectrum). The second is to compute graph features using existing methods. Other approaches are described in the main text. (d) Using feature vectors and labels, a classification model is trained, validated and tested.}
\label{fig:Diagram}
\end{figure}

Here, we briefly review Koopman spectral analysis, which is the underlying theory for various DMDs, and then describe the Graph DMD framework.
First, we consider a nonlinear dynamical system: $\bm{x}_{t+1} = \bm{f}(\bm{x}_t)$, where $\bm{x}_t$ is the state vector in the state space $\mathcal{M} \subset \R^{p}$ with time index $t\in\mathbb{T}:=\mathbb{N}_0$ .
{\em The Koopman operator}, which we denote by ${\cal K}$, is a linear operator acting on a scalar observable function $g\colon {\cal M}\to \C$ defined by
\begin{equation}\label{eq:koopman}
{\cal{K}}g=g\circ\bm{f},
\end{equation}
where $g\circ\bm{f}$ denotes the composition of $g$ with $\bm{f}$ \cite{Koopman31}.
That is, it maps $g$ to the new function $g\circ\bm{f}$.
We assume that ${\cal K}$ has only discrete spectra.
Then, it generally performs an eigenvalue decomposition:
$\mathcal{K}{\varphi}_{j}(\bm{x})={\lambda}_{j}{\varphi}_{j}(\bm{x})$,
where $\lambda_{j}\in\C$ is the \textit{j}-th eigenvalue 
(called \textit{the Koopman eigenvalue}) and $\varphi_{j}$ is 
the corresponding eigenfunction (called \textit{the Koopman 
eigenfunction}).
We denote the concatenation of scalar functions as $\bm{g} := [g_{1},\ldots, g_{d}]^{\mathsf{T}}$.
If each $g_{i}$ lies within the space spanned by the eigenfunction $\varphi_{j}$, we 
can expand the vector-valued $\bm g$ in terms of these 
eigenfunctions as $\bm g (\bm{x})=\sum_{j=1}^{\infty }{\varphi_{j}(\bm{x})\bm{\psi}_{j}}$, 
where $\bm{\psi}_{j}$ is a set of vector 
coefficients called \textit{the Koopman modes}. Through the iterative applications of $\mathcal{K}$, the following equation is obtained:
\begin{flalign}\label{eq:decomposition}
\bm{g}(\bm{x}_t)
=(\bm{g}\circ \underbrace{\bm{f}\circ\cdots\circ\bm{f}}_t)\left(\bm{x}_0\right)
=\sum_{j=1}^{\infty}{\lambda}_j^t{\varphi}_j\left(\bm{x}_0\right)\bm{\psi}_j.
\end{flalign}
Therefore, $\lambda_{j}$ characterises the time evolution of the corresponding Koopman mode $\bm{\psi}_j$, 
i.e. the phase of $\lambda_{j}$ determines its frequency and 
the magnitude determines the growth rate of its dynamics.

Among several possible methods to compute the above modal decomposition from data, DMD \cite{Rowley09,Schmid10} is the most popular algorithm, which estimates an approximations of the decomposition in Eq. (\ref{eq:decomposition}).
For the details of basic DMD and its variants, see Material and Methods.
Among several variants of DMDs, Graph DMD\cite{Fujii19b} can extract and visualise the underlying low-dimensional global dynamics of GDSs with structures among observables from data.
Then, we briefly introduce Graph DMD framework, described in Figs. \ref{fig:Diagram}a and b.
Here, we consider an autonomous discrete-time weighted and undirected GDS defined as
\begin{flalign}\label{eq:GDS}
G = (\mathcal{V},\mathcal{E},\bm{x}_t,\bm{y}_t,\bm{f}, \bm{g}_{\bm{c}}, \bm{A}_t),
\end{flalign}
where $\mathcal{V}=\{V^1,\ldots,V^m\}$ and $\mathcal{E} = \{E^1,\ldots,E^{l}\}$ are the vertex and edge sets of a graph, respectively, fixed at each time $t\in\mathbb{T}:=\mathbb{N}_0$.
$\bm{x}_{t}$ is the state vector in a state space ${\cal M} \subset \R^{p}$ for the GDS and $\bm{f}\colon\cal{M}\to\cal{M}$ is a (typically, nonlinear) state-transition function (again, $\bm{x}_{t+1} = \bm{f}(\bm{x}_t)$).
$\bm{y}_{t} = [\bm{y}_{1,t}^\mathsf{T},\ldots,\bm{y}_{m,t}^\mathsf{T}]^\mathsf{T}$ are concatenated observed values, where $\bm{y}_{i,t} \in \R^d$ are observed values for a vertex $i = 1,\ldots,m$.
They are given by $\bm{y}_{t}:= \bm{g}_{\bm{c}}(\bm{x}_t)$, where $\bm{g}_{\bm{c}} = [\bm{g}_{1}^\mathsf{T},\ldots,\bm{g}_{m}^\mathsf{T}]^\mathsf{T}$ is a concatenated vector-valued observable function (i.e. $\bm{g}_{\bm{c}}\colon \mathcal M\to \R^{md}$ ).
$\bm{A}_t \in \R^{m \times m}$ is an adjacency matrix, whose component $a_{i,j,t}$ represents the weight on the edge between $V^i$ and $V^j$ at each time $t$.
For example, the weight represents the relation (e.g. a function of distance) between moving agents in multi-agent systems.

For the formulation of Graph DMD, we propose a more straightforward formulation with dependent structure among observables than the previous formulation based on vector-valued reproducing kernel Hilbert spaces (RKHSs)\cite{Fujii19b}. 
Regarding the details of the analysis, the connection with Graph DMD (such as the relation between $\bm{A}_t$ and $\bm{g}_{\bm{c}}$) and the reason why it can visualise the relation between elements, see Materials and Methods. 
For the practical implementation of the spectral decomposition, a modified tensor-based DMD \cite{Fujii19b,Klus18}, which is a generalised DMD for the application to tensor data, is applied to the adjacency matrix series.

In Graph DMD, a sequence of the matrices $\bm{A}_t$ or an order-3 tensor ${\mathcal A}_{:,:,t}$ for $t = 0,\ldots,\tau$ is given, where colons are used to indicate all elements. 
In a similar way of basic DMD procedure (see Materials and Methods), we define tensors $\mathcal{X}, \mathcal{Y}$ such that $\mathcal{X}_{:,:,t} = \mathcal{A}_{:,:,t}$ and $\mathcal{Y}_{:,:,t} = \mathcal{A}_{:,:,t+1}$ for $t = 0,\ldots,\tau-1$.
Instead of singular value decomposition (SVD) in basic DMD, Graph DMD utilises tensor-train decomposition\cite{Oseledets11} to maintain the tensor structure of input data.
Here, we compute $\bm M \in \R^{m^2 \times r_2}$, $\bm{\Sigma} \in \R^{r_2 \times r_2}$ and $\bm N \in \R^{r_2 \times \tau}$ by matricising after tensor-train decomposition of $\mathcal X$ ($r_2$ is a tensor-train rank in the decomposition and $\bm{\Sigma}$ is a full-rank diagonal matrix; for further understanding, see the previous paper\cite{Fujii19b}). 
Similarly, we compute $\bm P \in \R^{m^2 \times s_2}$ and $\bm Q \in \R^{s_2 \times \tau}$ by matricising after tensor-train decomposition of $\mathcal Y$, where $s_2$ is a tensor-train rank. 
Note that this is similar to SVD in the matrix form, but SVD and this matrisation after tensor-train decomposition with maintaining the tensor structure are completely different. 
After that, in a similar way of basic DMD procedure, we then define a matrix $\hat{\bm F} = (\bm M^* \cdot \bm P)(\bm Q \cdot \bm N^\dagger)\bm{\Sigma}^{-1}$, where $\bm M^*$ is the Hermitian transpose of $\bm M$ and $\bm N^\dagger$ is the pseudo-inverse of $\bm N$.
Thereafter, we perform eigendecomposition of $\hat{\bm F}$ and obtain eigenvectors $\bm{w}_j$ and eigenvalues $\hat{\lambda}_j$ for $j = 1,\ldots,p$.
The latter is the estimated Koopman eigenvalues called {\it Graph DMD eigenvalues}.
Finally, we obtain the spatial coefficients $\bm{Z}_j \in \C^{m \times m}$ by matricising $\bm{z}_j = (1/\hat{\lambda}_j)\cdot \bm{P} \, \bm{Q} \cdot \bm{N}^\dagger \, \bm{\Sigma}^{-1} \cdot \bm{w}_j$ in the inputted adjacency matrix form, which are called {\it Graph DMD modes}.
In summary, a sequence of adjacency matrices is decomposed into spatial coefficients and temporal dynamics (Fig. \ref{fig:Diagram}b):
\begin{equation}\label{eq:estimatedGDMD}
\bm{A}_t \approx \sum^p_{j=1}\bm{Z}_j\hat{\lambda}_j^t {b}_{j,0},
\end{equation}
where ${b}_{j,0}$ works as an initial value described in Materials and Methods (we also describe other detailed procedures there).
 
\subsection*{Examples of Graph DMD for ballgame data.}

\begin{figure}[!t]
\centerline{\includegraphics[width=1 \textwidth]{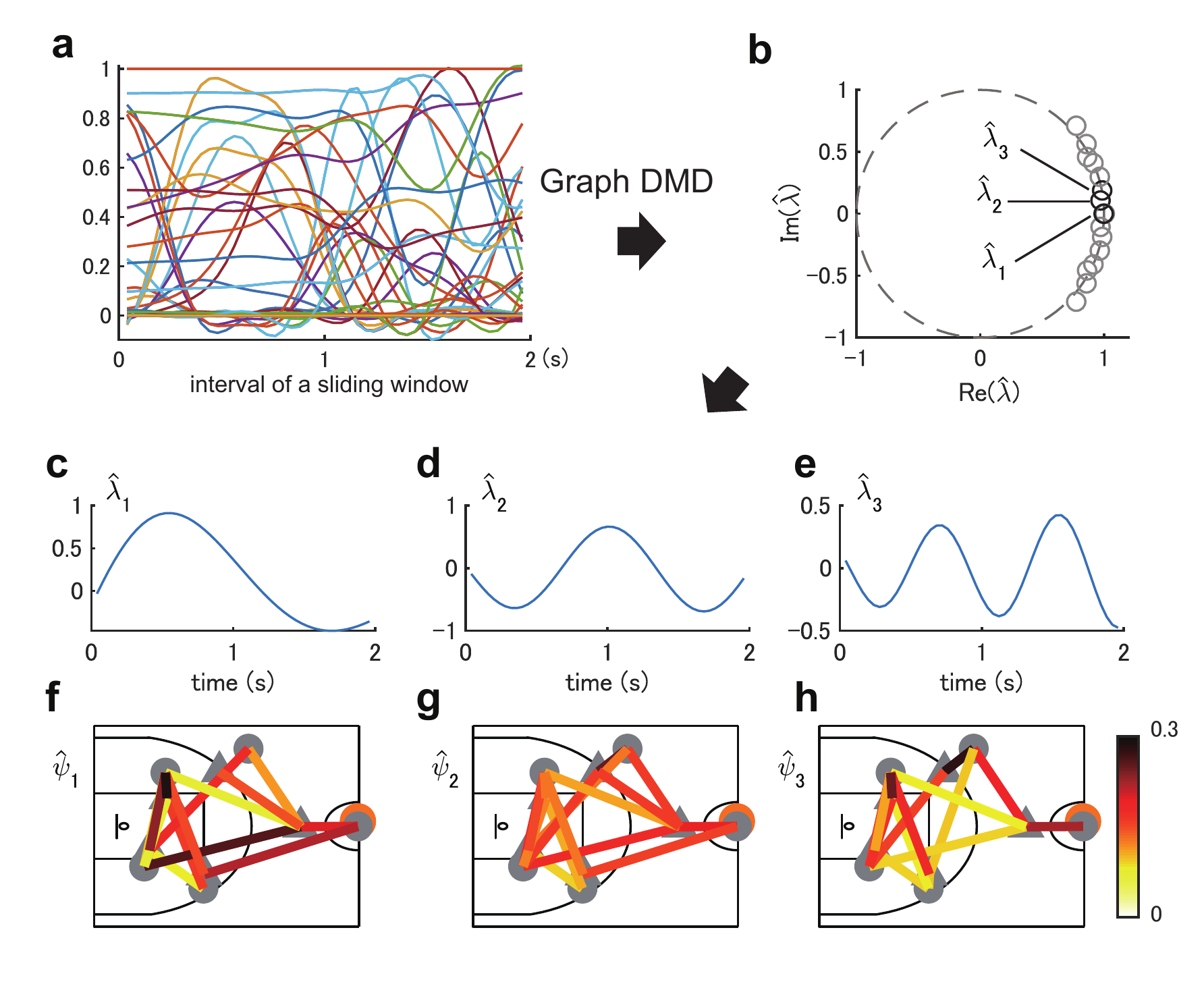}}
\caption{{\bf A representative example of Graph DMD.}
(a) An example of obtained data (components of adjacency matrix series). 
(b) Graph DMD eigenvalues (black circles) computed by Graph DMD in a complex plane.
(c-e) Time dynamics and (f-h) visualised Graph DMD modes for the dynamics with the lowest frequencies.
}
\label{fig:example}
\end{figure}

Next, we show a representative example of Graph DMD result in Fig. \ref{fig:example}.
Here we used player-tracking data from actual basketball games. 
Position data was composed of the horizontal Cartesian positions of every player and the ball on the court, recorded at 25 frames per second.  
We analysed an attack-segment defined as the period that begins when all players enter the attacking side of the court and ends before a shot.
The input of Graph DMD is adjacency matrix series $\bm{A}_t$ visualised in Fig. \ref{fig:Diagram}a and its elements are:

\begin{equation} \label{eq:GaussKernel}
  A_{i,j,t} = \exp\left(\frac{-\|\bm{y}_{i,t}-\bm{y}_{j,t}\|^2_2}{2\sigma}\right),
\end{equation} 
where $i$ and $j$ indicate players and a ring (i.e. a goal as geometric information), $\|\cdot\|_2$ is the Euclidean norm and $\sigma$ is a coefficient for adjusting the value of $A_{i,j,t}$, which represents a proximity with contextual meaning in this case (for details, see Materials and Methods).  
Furthermore, since the order of $i$ is not uniquely determined in general, we sorted the adjacency matrix in order of nearest attackers and defenders from the ball in each time stamp, to provide the order of player to the semantic information. 
Here, we denote the first to the fifth nearest attackers and defenders to the ball as A1, ..., A5 ($i = 1,\ldots,5$) and D1 $,\ldots,$ D5 ($i = 6,\ldots,10$), respectively (for the ring, $i = 11$). 
Also, we denote the relationship between two players, e.g. A1 and D1 as A1-D1. 

Firstly, we obtained DMD eigenvalues via Graph DMD. 
As illustrated in Fig. \ref{fig:example}b, all eigenvalues appear to be on the unit circle.
Eigenvalue $\hat{\lambda}_j$ is transformed into temporal frequency $\omega_j$ such that $\omega_j = \ln (\hat{\lambda}_j) /2\pi \Delta t$, where $\Delta t$ is a time interval of the discrete time system (i.e. $\Delta t = 1/25$).
Then, for each eigenvalue, we obtained the time dynamics in Fig. \ref{fig:example}c-e and Graph DMD modes in Fig. \ref{fig:example}f-h. 
For the sake of clarity, only three pairs of the dynamics and modes showing the smallest frequency in this order are presented.
Graph DMD extracted the dynamics with approximately one, two and three cycles.
For the Graph DMD modes, we visualised the strength of the spectrum (i.e. coefficient) for each time dynamics.
Note that DMD computes complex-valued DMD modes and time dynamics, but here we only present the absolute value and real parts, respectively.
In Supplementary Text 2, we describe the selection of parameters for Graph DMD (e.g. tolerance, temporal window size and cutoff frequency) and quantitatively validated the applicability of Graph DMD to the sport data in terms of the reconstruction error.
 
\subsection*{Classification in various global collective behaviours.}
\begin{figure}[t!]
\centerline{\includegraphics[width=1 \textwidth]{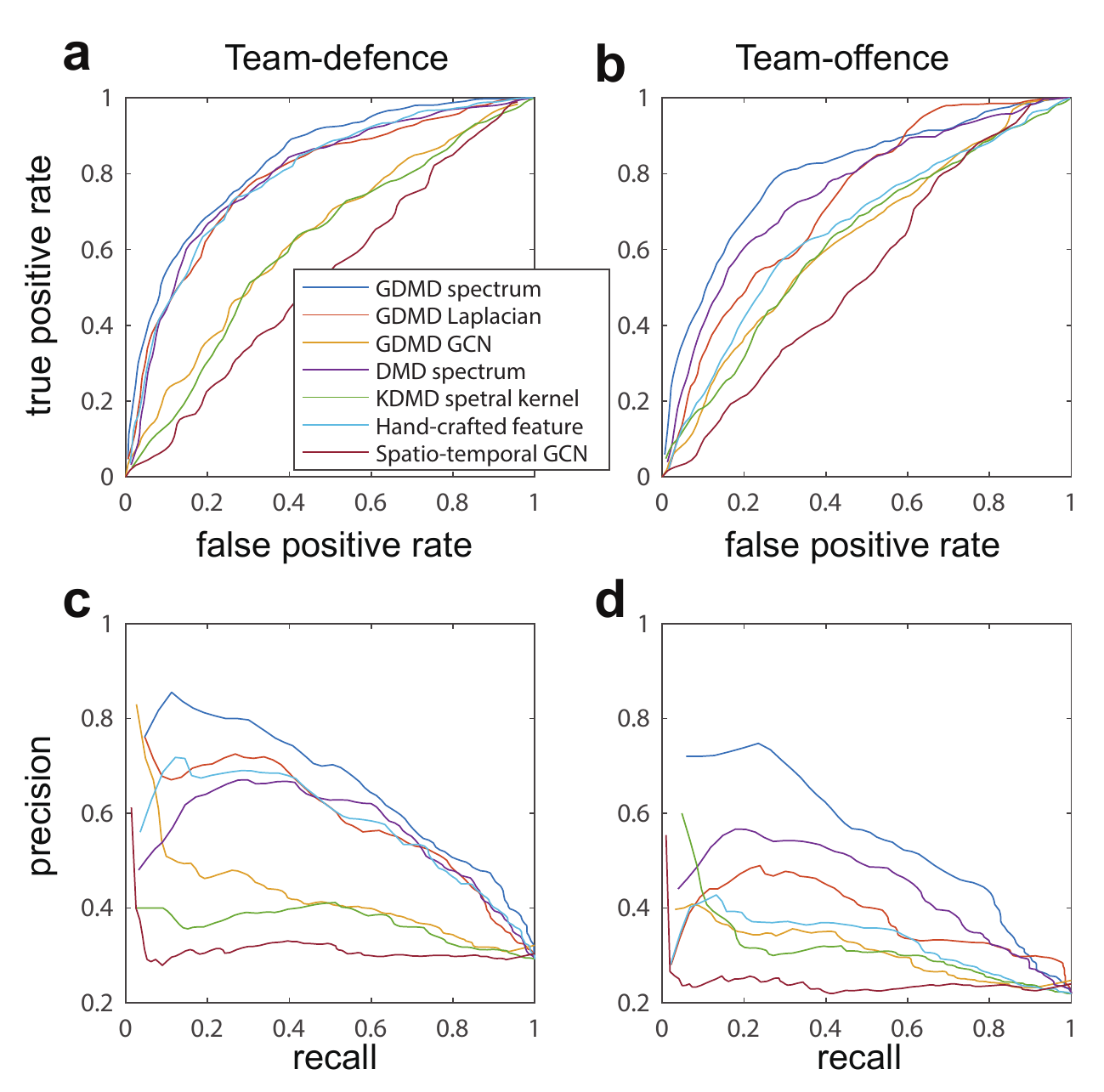}}
\caption{{\bf Classification results of seven methods in two recognition tasks.} Receiver operating characteristic (ROC) curves in (a) the team-defence and (b) team-offence recognition task are shown. Also, trade-off curves between recall and precision in (c) the team-defence and (d) team-offence recognition task are shown. The seven methods are given in the legend of (a) and the text.}
\label{fig:classify}
\end{figure}

Next, we indicate that our methods had better classification performance than most of existing methods in Table \ref{tab:classify} and Fig. \ref{fig:classify}.
As validation, we performed two classification tasks with different global collective behaviours: the team-defence (zone or person-to-person defence) and team-offence (offence with and without screen-play) recognition tasks. 
By definition, zone and person-to-person defence are exclusive (i.e. players guard their area and opponent players, respectively). 
However, they are actually mixed based on different situations whereas experienced people can distinguish them. 
A screen-play is the basic and minimal strategic cooperative play in basketball\cite{Hojo18}, in which an attacker stands on the course of defence player like a 'screen' and prevents the defender from defending another attacker in a legal way. 
Additional details are provided in Materials and Methods.

After Graph DMD, we performed classification using three different approaches.
The first two approaches create feature vectors and the third automatically extracts the features via a neural network approach.
The first approach is simply to vectorise the Graph DMD modes (denoted GDMD spectrum) to directly reflect the node (i.e. player) information. 
For the team-defence recognition task, we used the elements of the modes regarding the relations among defenders, those among attackers and defenders and those among defenders and ring (as geometric information) as a feature vector.
For the team-offence recognition task, we used those among attackers and among attackers and defenders (without geometric information).
Selection and elimination were based on prior knowledge of the sport.
The final two approaches computed graph features using existing methods: the second is graph Laplacian eigenvalues \cite{Chung97} as a baseline feature of a graph \cite{deLara18} (denoted GDMD Laplacian) and the third is deep graph convolutional neural networks \cite{Zhang18} (denoted GDMD GCN) as a recent promising method for classifying the graph data.
The second approach extracts the graph topology without node information, and the third one retains more node information and learns the graph topology.

Due to the highly-strategic (or contextual) behaviours in this experiment, we hypothesised that, in addition to physical information, the contextual node information (i.e. the nodes themselves in the first approach) is more important for the classification than the graph features (i.e. the second and third approaches).
For all classification methods with the exception of the neural network approaches, we adopted logistic regression as a simple linear binary classification model.

As comparable methods for classifying this data, we adopted four methods.
The first is a method using vectorised basic DMD modes \cite{Tu14} (denoted DMD spectrum) as a baseline of DMD approaches.
The second is the Koopman spectral kernel\cite{Fujii17} using DMD with reproducing kernels\cite{Kawahara16} as the existing method for classifying the collective motion dynamics\cite{Fujii18} (denoted KDMD spectral kernel). 
For these two methods, the input data must be a matrix; thus we did not utilise the structure of the graph sequence data.
The third is a hand-crafted feature as a simple baseline method, consisting of the vectorised temporal average, maximum and minimum values of the elements of the input adjacency matrix series.
Fourth is an advanced neural network approach for classifying graph sequence data called spatio-temporal GCN \cite{Yan18} as a recent promising end-to-end method.
More information on the selection and the details of these methods are provided in Supplementary Text 1.

We investigated classification performance in Table \ref{tab:classify} in terms of accuracy, the area under the curve (AUC) based on receiver operating characteristic (ROC) curve in Fig. \ref{fig:classify}a and b, and F-measure, which is the trade-off between recall and precision (the curve is shown in Fig. \ref{fig:classify}c and d).
Overall, the classification performance in GDMD spectrum was better than those in most of the remaining methods.
In the statistical evaluation, there were significant differences in all classification performance and tasks ($p > 3.63 \times 10^{-8}$) using analysis of variance (ANOVA) or Friedman test.
In the following evaluations, we indicate the post-hoc comparison results.
Accuracy in GDMD spectrum was significantly higher than that of other methods ($r > 0.39$ and $ p < 0.03$) for both recognition tasks, with the exception of GDMD Laplacian and DMD spectrum in the team-defence recognition ($ r > 0.16$ and $ p > 0.05$).
AUC and F-measure in GDMD spectrum were also higher than those of the other methods ($r > 0.40$ and $ p < 0.028$) for both recognition tasks, with the exception of hand-crafted feature in the team-defence recognition (both AUC and F-measure: $ r > 0.34$ and $ p > 0.05$).
As a feature extraction method, Graph DMD was a better method than most of the other methods, and especially simple Graph DMD spectrum was better than the graph features in the collective motions. 

\newcommand{\md}[2]{\multicolumn{#1}{c|}{#2}}
\newcommand{\me}[2]{\multicolumn{#1}{c}{#2}}
\begin{table}[ht!]
\centering
\scalebox{0.7}{
\begin{tabular}{|l|rrr|rrr|}
\hline
& \md{3}{Team-defence} & \md{3}{Team-offence} \\
& \me{1}{Acc} & \me{1}{AUC} & \md{1}{F-measure} & \me{1}{Acc} & \me{1}{AUC} & \md{1}{F-measure} \\ 
\hline
GDMD spectrum & $0.785 \pm0.062 $ & $0.833 \pm0.057$ & $0.533 \pm0.058$ & $0.809 \pm0.046 $ & $0.803 \pm0.061$ & $0.455 \pm0.060$ \\GDMD Laplacian & $0.756 \pm0.045 $ & $0.787 \pm0.062$ & $0.506 \pm0.057$ & $0.770 \pm0.047 $ & $0.730 \pm0.057$ & $0.405 \pm0.057$ \\GDMD GCN & $0.708 \pm0.040 $ & $0.678 \pm0.076$ & $0.406 \pm0.078$ & $0.782 \pm0.058 $ & $0.626 \pm0.110$ & $0.338 \pm0.069$ \\
\hline
DMD spectrum & $0.765 \pm0.051 $ & $0.793 \pm0.068$ & $0.509 \pm0.055$ & $0.794 \pm0.046 $ & $0.755 \pm0.062$ & $0.425 \pm0.056$ \\KDMD spetral kernel & $0.664 \pm0.052 $ & $0.612 \pm0.094$ & $0.407 \pm0.074$ & $0.736 \pm0.067 $ & $0.625 \pm0.103$ & $0.350 \pm0.069$ \\Hand-crafted feature & $0.756 \pm0.050 $ & $0.793 \pm0.040$ & $0.510 \pm0.047$ & $0.745 \pm0.048 $ & $0.655 \pm0.076$ & $0.369 \pm0.054$ \\Spatio-temporal GCN & $0.711 \pm0.104 $ & $0.528 \pm0.086$ & $0.338 \pm0.091$ & $0.782 \pm0.051 $ & $0.540 \pm0.077$ & $0.289 \pm0.067$ \\
\hline
\end{tabular}}
\caption{\label{tab:classify}{\bf Classification performance of seven methods for two recognition tasks.} {\rm Accuracy (Acc), area under the curve (AUC) based on receiver operating characteristic (ROC) curve and F-measure are indicated.  Overall, classification performances in GDMD spectrum were higher than those in most of remaining methods.}}
\end{table}
 

\subsection*{Analysis of visualised Graph DMD modes.}
Since our method can decompose the data into temporal dynamics and spatial coherent structures in the observed data space, we can interpret the decomposed modes that contribute to the classification results. 
Here we show the averaged spectra for each label and classification task to identify the trends (note that the location of collective motion networks shown in Fig. \ref{fig:example} change with time; thus for an accurate understanding, we averaged and visualised them in an adjacency matrix form). 
 
\begin{figure}[t!]
\centerline{\includegraphics[width=1 \textwidth]{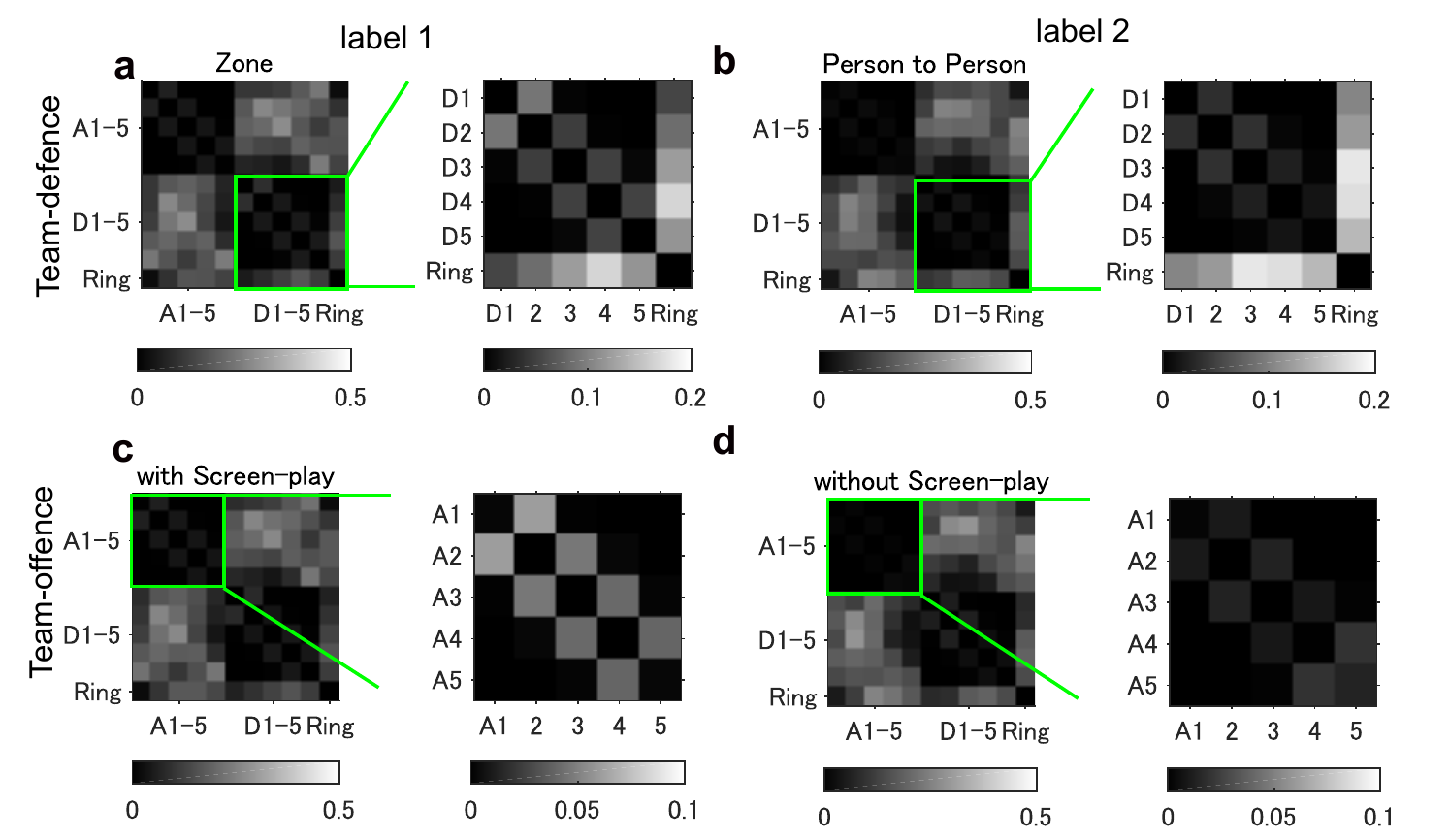}}
\caption{{\bf Averaged Graph DMD modes.}
Averaged Graph DMD modes for label 1 (a and c) and 2 (b and d) in the team-defence (a and b) and offence (c and d) recognition tasks are shown. 
We denote the first to fifth nearest attacker and defender as A1 $,\ldots,$ A5 and D1 $,\ldots,$ D5, respectively. 
R indicates the ring (goal).
A strong spectrum for (a) zone defence was observed in the relationship among the nearest defenders; however, it was not observed in (b) person-to-person defence.
Similarly, a strong spectrum for (c) offence with screen-play was observed in the relationship between the nearest attackers; however, it was not observed for (d) offence without screen-play.
}
\label{fig:GDMDmode}
\end{figure}

\begin{figure}[t!]
\centerline{\includegraphics[width=1 \textwidth]{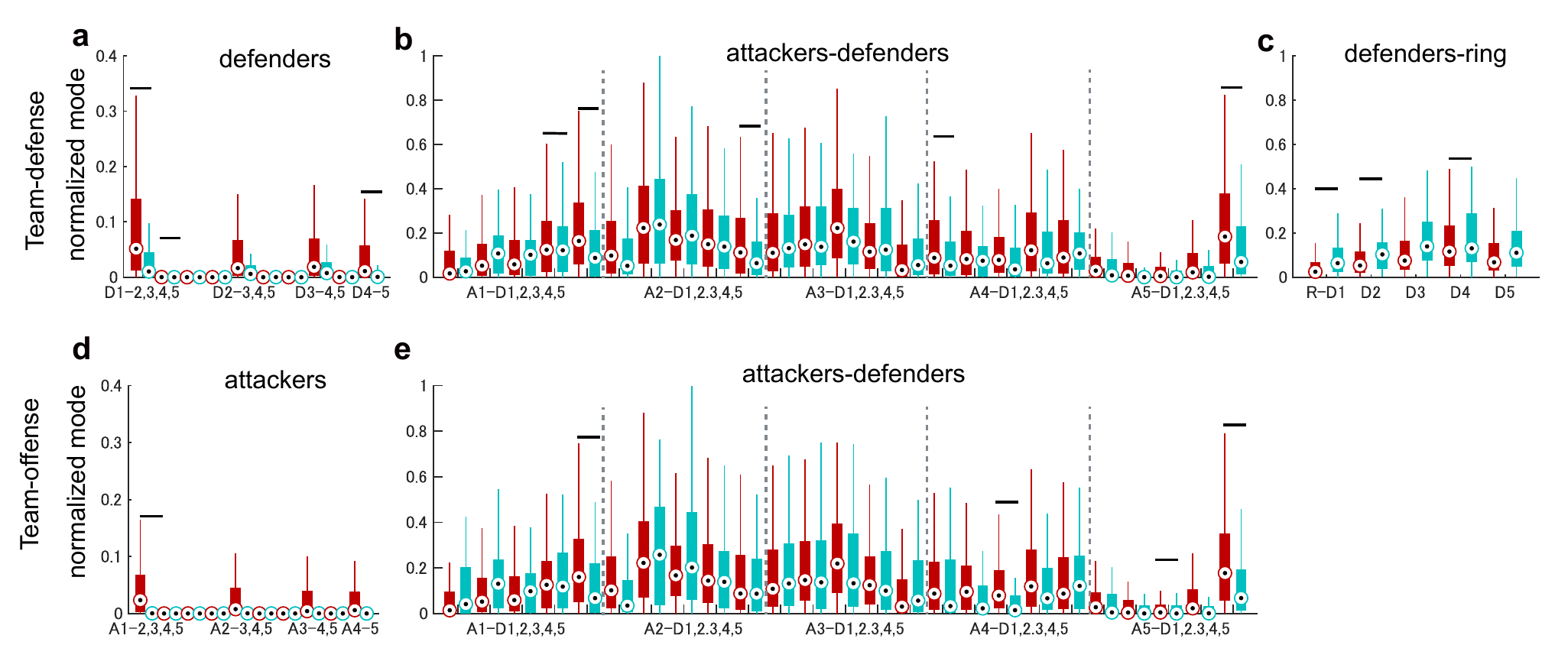}}
\caption{{\bf Boxplots of feature vectors as elements of Graph DMD modes.}
Feature vectors as elements of Graph DMD modes for label 1 (red) and 2 (green) regarding the feature vectors used in the team-defence (a-c) and offence (d and e) recognition tasks are shown (label 1: zone defence and offence with screen-play; label 2: person-to-person defence and offence without screen-play).
For the team-defence recognition task, the elements of GDMD modes (a) among defenders, (b) those among attackers and defenders and (c) those among defenders and the ring were selected.
For the team-offence recognition task, those (d) among attackers and (e) those among attackers and defenders were selected.
Notations are same as Fig. \ref{fig:GDMDmode}.
Horizontal bars indicate that the relationship is significantly explained by logistic regression.
For all statistical results, see Supplementary Tables 1 and 2.}
\label{fig:Boxplot}
\end{figure}

For the team-defence recognition task, a strong spectrum for zone defence in Fig. \ref{fig:GDMDmode}a was observed in the relationship among the nearest defenders (expanded in the figure) but it was not observed in person-to-person defence Fig. \ref{fig:GDMDmode}b.
The variability of the feature vectors is also presented via boxplots in Figs. \ref{fig:Boxplot}a-c.
In examining the individual contribution, statistical analysis using logistic regression indicates that the relationships among defenders (D1-D2 and D4-D5: odds ratio $> 31.81$ and $p < 0.023$; D1-D3 was also significant but too small), those among attackers and defenders (A1-D4, A1-D5, A2-D5, A4-D1, and A5-D5: odds ratio $> 0.02$ and $p < 0.023$) and those among defenders and ring (ring-D1, D2 and D4: odds ratio $> 0.002$ and $p < 0.049$) significantly explained the labels of team-defences. 
Note that this analysis separately investigated the individual contribution as well as the above classifications.
For more detailed statistical results and boxplots of other elements of DMD modes, see Supplementary Table 1 and Supplementary Fig. 2, respectively.

For the team-offence recognition task, a strong spectrum for offence with screen-play in Fig. \ref{fig:GDMDmode}c was observed in the relationship between the nearest attackers; however, it was not observed for offence without screen-play Fig. \ref{fig:GDMDmode}d (the boxplot is shown in Figs. \ref{fig:Boxplot}d and e).
Results of logistic regression indicate that the relationship in D1-D2 (odds ratio $= 1.06 \times 10^4 $ and $p = 5.36 \times 10^{-5}$) and those among attackers and defenders (A1-D5, A4-D3 and A5-D5: odds ratio $> 23.59$ and $p < 0.009$; A5-D3 was also significant but too small) significantly explained the labels of team-offence.
For more details, see Supplementary Table 2 and Supplementary Fig. 2.
With these methods, we can interpret the types of interaction between agents (and environments) that contributes to focusing dynamic properties (i.e. frequency in this case) with physical meaning. %
In Discussion section, we discuss further semantic interpretation.
 
\section*{Discussion}
The objective of this paper was to understand the network dynamics for complex collective motions via physically-interpretable classification using data-driven spectral analysis of GDSs called Graph DMD. 
In this section, we discuss the comparison with the conventional approaches, semantically interpret the visualised Graph DMD modes, describe the limitation and the future perspectives for our methods. 
We then present our conclusions.

Using ballgame data as an example, we classified the strategic collective motions in the team-defence and offence recognition tasks more successfully than most of the existing methods.
The results of the comparison with graph features (GDMD Laplacian and GCN) suggest that the collective motions in this case can be classified with the node information (nearest to the ball) rather than the graph features.
The results of other DMD methods (DMD spectrum and KDMD spectral kernel) suggest that both the reflecting graph structure and the use of appropriate sliding windows were important for recognition. 
Furthermore, in comparison with KDMD spectral kernel \cite{Fujii17} (i.e. decomposition in a feature space), the proposed method has an advantage in physically and semantically interpreting the DMD modes in the observed data space.
Regarding an end-to-end neural network approach (i.e. spatio-temporal GCN), the model for motion capture data (i.e. joint location data for a few people) must be customised.
However, each agent is not physically connected in this case; thus, the positional relationship (i.e. constraint condition) dynamically changes in contrast to the motion capture data. 
In this paper, we used a simple setting with minimal correction; however, further adaptation to complex collective motions is required (e.g. more appropriate customisation of the network and adequate learning such as using a larger amount of data).
For hand-crafted features, we compute simple features to demonstrate the validity of our methods in this paper.
The use of a more customised framework\cite{Hojo18} which includes the detection of a few related people might improve specific classification performance.
However, in this paper, we proposed a more generalised classification framework for global dynamics of collective motions, which successfully performed two different recognition tasks (team-defence and offence) by selecting only the elements of the Graph DMD modes as feature vectors.
Furthermore, our method can obtain the dynamical properties of the collective motion network.

The proposed method can obtain physically-interpretable dynamical properties of network dynamics and can perform classification even for complex collective motions.
For example, in Fig. \ref{fig:Boxplot}, we found the label-specific stronger spectra in two recognition tasks, in which the relationships among the nearest attackers or defenders, those among attackers and defenders far from the ball and those between defenders and the ring, which were oscillated in low-frequency modes.
Semantically, in zone defence, defenders adjusted their positions more according to a teammate (Fig. \ref{fig:Boxplot}a), the attacker with the ball (Fig. \ref{fig:Boxplot}b: especially in the distant case from the ball) and less according to the geometric reference point (Fig. \ref{fig:Boxplot}c) than the defenders in person-to-person defence.
For team-offence with screen-play, attackers adjusted their positions more according to a teammate (Fig. \ref{fig:Boxplot}d) and the distant defender from the ball (Fig. \ref{fig:Boxplot}e) than the attackers in team-offence without screen-play. 
Generally, our approach can be applied to the analysis of complex global dynamics in groups of living organisms or artificial agents, which currently eludes mathematical formulation.  
The human group in a sport used in this study can be considered as an examples in which communication is likely to be measured in a physical space \cite{Fujii18}. 
Therefore, by measuring the outputs of communication and assuming that there are underlying dynamics behind the obtained data, our approach can handle various means of communication between agents.
In practice, in various sports teams or other human communities, supervisors (e.g. experimenters, coaches and teachers) spend considerable amounts of time analysing the collective motions in their domain. Application of a system, such as the one presented here, can lead to the creation of useful plans that are currently derived only from their implicit experience. 

However, there are some limitations to this study. 
One is the validation of whether our approach can extract true dynamical properties if used as an equation-free method, which cannot confirm the true dynamical properties (e.g. frequencies with growth/decay rate), as well as in the previous works \cite{Brunton16a,Proctor15}.
Originally, DMD demonstrates its strength for the dynamical systems which can be mathematically defined \cite{Kawahara16,Takeishi17c} or of which solutions are empirically known\cite{Rowley09,Schmid10}. 
We instead validated our approach using classification performance, a qualitative evaluation with specific knowledge of the sport domain (e.g. low-frequency band) and the reconstruction error (see Materials and Methods).
However, a general quantitative validation method for the unknown dynamics is further required.
Another is to reflect the more local interaction dynamics such as local competitive and cooperative play by the attackers and defenders \cite{Fujii14,Fujii15,Fujii15b,Fujii16}, which can provides more practical information in the sport domain.
Although the purpose of Graph DMD is to extract the underlying global dynamics of GDSs and we can obtain the interpretable local spectra in the DMD modes, there are other approaches for extracting the more specific local dynamics.
Even when using only players' location data such as in this study, more specific methods reflecting local competition can be applied to more practical application such as score prediction \cite{Fujii17,Fujii18} and prediction of a player to obtain the ball after shot \cite{Hojo19}.

In conclusion, we applied a data-driven spectral analysis called Graph DMD, which obtains physically-interpretable dynamical properties of network dynamics for the classification even for complex collective motions.
We classified the motions in different global behaviours and discovered that, in addition to the physical properties, the contextual node information was critical for classification.
Furthermore, we discovered the label-specific stronger spectra in the relationship between the nearest agents, which provided physical and semantic interpretation. 
Our approach contributes to the understanding of complex networks involving collective motions from the perspective of nonlinear GDSs.

\section*{Methods} 
\subsection*{Positional data in a ballgame.}
The positional data of players and the ball (25 frames per second) were obtained from actual men's Asian international level practical games held in 2015 and preprocessed by STATS SportVU system (Northbrook, IL, USA). We obtained the consent to use it for research. We analysed 220 min of play (in 4 days) in which the two teams scored 746 points (386 vs. 360). For each day, players performed one and a half games (i.e. 60 min) except for 1 day (only one game). The positional data contained the XY position of each player and the XYZ coordinates of the ball on the court (All-court: $28 \times 15$m; half-court: $14 \times 15$ m). After the following data segmentation, we obtained 319 attack-segments.

\subsection*{Data segmentation.}
Prior to data segmentation, we used a custom-made automatic individual play-detection system to detect shots using positional data similar to that used in previous studies \cite{Fujii16,Fujii18,Hojo18}. 
We analysed an attack-segment defined as the period that begins when all players enter the attacking side of the court and ends before a shot (we analysed only the attack-segment finishing with a shot).
Then, the authors, who have experience in playing and coaching basketball, manually labelled all attack-segments into a zone defence, a person-to-person defence as the team-defence recognition task and offence with and without screen-plays as the team-offence recognition task.

Person-to-person defence is a team-defence strategy in which defenders basically guard their predetermined attackers.
In contrast, zone defence is another team-defence strategy in which defenders basically guard their predetermined areas.
Although zone and person-to-person defence are exclusive by definition, they are actually mixed according to different situations.
For example, against an attacker's penetration with the ball to the ring, the defenders in person-to-person defence also guard the area near the ring; in contrast, against an attacker's shot, the defenders in zone defence also guard the attacker with the ball.
Despite this ambiguity, people with the experience of games can distinguish the two team-defence strategies by observing the defensive motion without the ball.

Screen-play is the basic and minimal strategic cooperative play in basketball\cite{Hojo18}, in which an attacker stands on the course of defence player like a 'screen' and prevents the defender from defending another attacker in a legal way. 
We labelled attack-segments into an offence with (at least including a screen-play) and without screen-play.
Note that we focused on the global network dynamics of collective motions, thus we did not segment and label screen-plays themselves (i.e. in a spatio-temporally accurate sense such as in the previous study\cite{Hojo18}).

We randomly created a validation dataset (30 attack-segments) for the selection of parameters regarding Graph DMD, and a test dataset (289 attack-segments) for the subsequent analyses.

\subsection*{Creating adjacency matrix series.}
Next, we compute the adjacency matrix sequence in the attack-segment for inputting to the subsequent graph DMD.
Here, using the Gaussian kernel in Eq. (\ref{eq:GaussKernel}), we converted the distances between all individuals and the geometric reference position (i.e. the ring) into weights of adjacency matrix series, which indicate 1 if close to each other and 0 if far from each other.
In this way, we can obtain more stable time series than those directly using the distances.
The numerator in the exponential in Eq. (\ref{eq:GaussKernel}) represents the distance between all individuals and the ring.
Since the order of $i$ cannot be uniquely determined in general, we must define the order based on reasonable grounds.
For example, the order of playing positions (e.g. guard, forward and centre) or jersey numbers does not provide fair comparison among attack-segments because the players themselves frequently changes throughout attack-segments.
A simple way to determine the order is based on the distance between players and a reference point (e.g. ball or ring).
However, if the order is simply determined by proximity to the reference point, there may arise a problem that, for example, the distance between the third and fifth players is shorter than that between those (the third and fifth) and the fourth players.
Thus, first, we simply set $i = 1,2$ for the two attackers in order from the attacker closer to the ball.
Then, we set $i=3$ for the closest attacker in terms of the sum of the distance to the above two attackers.
In this way, we set $i=4,5$ for the closest attacker based on the sum of the distance to the three and four attackers, respectively.
Similarly, we set $i = 6, \ldots, 10$ for the five defenders and set $i = 11$ for the ring position to take the geometric information into consideration.

Next, denominator $2\sigma$ in the exponential in Eq. (\ref{eq:GaussKernel}) is a coefficient for adjusting the value of $A_{i,j,t}$.
In this paper, we set $\sigma$ for the relationship between players that satisfies $A_{i,j,t} = 0.5$ when $\|\bm{y}_{i,t}-\bm{y}_{j,t}\|_2  = 1.5$ m (i.e. $\sigma = 1.5^2/2\rm{log}2$), which is considered to be a meaningful proximity between players (e.g. 1 m is near and 2 m is far in terms of the contact of two players).
Also, we set $\sigma$ for the relation between players and the ring that satisfies $A_{i,j,t} = 0.5$ when $\|\bm{y}_{i,t}-\bm{y}_{11,t}\|_2  = 6$ m (i.e. $\sigma = 6^2/2\rm{log}2$).
In this way, a sequence of matrix $\bm{A}_t$ with $A_{i,j,t}$ as an element at time $t$ is created as a sequence of adjacent matrices of the graph.
In this paper, we used this adjacency matrix series for both team-defence and offence recognition tasks. 

Selection of an appropriate representation of the data is a fundamental problem in pattern recognition.
Time-series data is challenging to design features for because of difficulties in reflecting the data structure (including time length). 
A simple method involves using the representative values (e.g. average and maximum values).
However, for time series data, which cannot be distinguished by these representative values, it is necessary to extract dynamic properties of the time series.
One of the effective methods is DMD\cite{Rowley09,Schmid10} as described below.

\subsection*{Basic DMD and its variants.}
\label{sec:koopman}

Here, we first describe the basic DMD procedure and briefly explain the variants of DMDs including Graph DMD.
Among several possible methods to compute the spectral decomposition from data, DMD \cite{Rowley09,Schmid10} is the most popular algorithm, which estimates an approximations of the decomposition in Eq. (\ref{eq:decomposition}). 
Consider a finite-length observation sequence $\bm{y}_0,\bm{y}_1,\ldots ,\bm{y}_{\tau}$ (${\in \C}^{m}$), where $\bm{y}_{ }:= {\bm{g}}(\bm{x}_t)$. 
Let $\bm{X}= [\bm{y}_0,\bm{y}_1 ,\ldots , \bm{y}_{\tau-1}]$ and $\bm{Y}= [\bm{y}_1,\bm{y}_2,\ldots , \bm{y}_{\tau}]$. 
Then, DMD basically approximates it by calculating the eigendecomposition of 
matrix $\bm{F}=\bm{Y}\bm{X}^{\dagger}$, where $\bm{X}^{\dagger}$ is the pseudo-inverse of $\bm{X}$. The matrix $\bm{F}$ may be intractable to analyse directly when the dimension is large. 
Therefore, in the popular implementation of DMD called exact DMD \cite{Tu14}, a rank-reduced representation $\hat{\bm F}$ based on SVD is applied.
That is, $\bm{X}\approx {\bm U}{\bm{\Sigma}}{\bm V}^*$ and 
$\hat{\bm F} = \bm U^*\bm F \bm U= \bm U^* \bm Y \bm V \bm{\Sigma}^{(-1)}$, where $^*$ is the conjugate transpose.
Thereafter, we perform eigendecomposition of $\hat{\bm F} \in \C^{p\times p}$ to obtain the set of the eigenvalues $\hat{\lambda}_j$ and eigenvectors $\bm{w}_j$.
Then, we estimate the Koopman modes in Eq. (\ref{eq:decomposition}):
$\bm{\psi}_j=\hat{\lambda}_j^{(-1)}\bm Y \bm V \bm \Sigma^{(-1)}\bm{w}_j \in \C^{m\times p}$, 
which is called \textit{DMD modes}.
Time dynamics of $j$th mode is defined as $\hat{\lambda}_j^t b_{j,0}$, where $b_{j,0} = \bm{\psi}_j^{\dagger}\bm{y}_0$  and $\bm{\psi}_j^{\dagger}$ is the $j$-th row of the pseudo-inverse of $[\bm{\psi}_1 ~ \cdots ~ \bm{\psi}_p]$\cite{Kutz16a}.
In summary, obtained time series data is decomposed into spatial coefficients and temporal dynamics: ${\bm y}_t \approx \sum^p_{j=1}\bm{\psi}_j\hat{\lambda}_j^t b_{j,0}$.

Theoretically, for DMD to compute the decomposition Eq.~\eqref{eq:decomposition}, each observable function $g_{i}$ should lie within the space spanned by the Koopman eigenfunction $\varphi_{j}$, i.e. the data should be rich enough to approximate the eigenfunctions.
However, in basic DMD algorithms naively using the obtained data such as exact DMD, the above assumption is not satisfied such as when the data dimension is too small to approximate the eigenfunctions.
Thus, there are several algorithmic variants of DMDs to overcome the problem of the original DMD such as a formulation in RKHSs\cite{Kawahara16}, in a multitask framework\cite{Fujii19a}, and using a neural network\cite{Takeishi17c,Lusch18}.
The previous Graph DMD\cite{Fujii19b} utilises the structure of graph sequences (i.e. a tensor structure of adjacency matrix series) by a formulation in a vector-valued RKHS and an implementation using tensor-train decomposition.
Although the previous Graph DMD is based on a general formulation in a vector-valued RKHS, in this paper, we propose a more straightforward formulation of Graph DMD in an observed data space.

\subsection*{Koopman spectral analysis with dependent structure among observables for Graph DMD.}

Here, we propose a more straightforward formulation for Graph DMD called {\it Koopman spectral analysis with dependent structure among observables} than the previous one\cite{Fujii19b}.
Notations are the same as those in Results.
The previous formulation of Graph DMD \cite{Fujii19b} was based on vector-valued RKHSs endowed with a symmetric positive semi-definite kernel matrix\cite{Alvarez12}.
Although the spectral decomposition of the vector-valued feature function in a vector-valued RKHS is generally formulated, in this study, we propose a more straightforward formulation connecting Koopman spectral analysis and Graph DMD (in Eq. (\ref{eq:estimatedGDMD})) in an observed data space (i.e. in an adjacency matrix form).
First, again, consider a vector-valued observable function $\bm{g}_{\bm{c}} = [\bm{g}_{1}^\mathsf{T},\ldots,\bm{g}_{m}^\mathsf{T}]^\mathsf{T}$.
To analyse a GDS with dependent structures among the elements of the observable function, $\bm{g}_1,\dots,\bm{g}_m$, we newly consider a matrix-valued observable function $\bm{G}\colon \mathcal M \to \R^{m\times m}$ as

\begin{equation}\label{eq:MVOF}
[\bm{G}(\bm{x})]_{i,j}=
h(\bm{g}_{i}(\bm{x}), \bm{g}_{j}(\bm{x})),
\end{equation}
for $i,j = 1,\ldots,m$, where $h$ is a function defined in the right-hand side of Eq.~\eqref{eq:GaussKernel}. 
Next, we consider a vector-valued function $\bm{g}_{\bm G}\colon \mathcal M \to \R^{m^2}$ defined as $\bm{g}_{\bm G}(\bm{x}) = \vectorize(\bm{G}(\bm{x}))$. %
In analogous ways of a basic Koopman spectral analysis (described in Results), the Koopman operator $\mathcal{K}_{\bm{G}}$ is defined by $\mathcal{K}_{\bm{G}}\bm{g}_{\bm G}=\bm{g}_{\bm G}\circ\bm{f}$, where $\bm{g}_{\bm G}\circ\bm{f}$ denotes the composition of $\bm{g}_{\bm G}$ with $\bm{f}$.
We assume that ${\cal K}_{\bm{G}}$ has only discrete spectra.
Then, it generally performs an eigenvalue decomposition:
$\mathcal{K}_{\bm{G}}{\varphi}_{j}(\bm{x})={\lambda}_{j}{\varphi}_{j}(\bm{x})$,
where $\lambda_{j}\in\C$ is the \textit{j}-th eigenvalue and $\varphi_{j}$ is the corresponding eigenfunction.
Note that the underlying nonlinear dynamical system is the same as the basic formulation, i.e. $\bm{x}_{t+1} = \bm{f}(\bm{x}_t)$ and $\bm{x}_t \in \mathcal{M} \subset \R^{p}$.
Then, if each element $[\bm{g}_{\bm G}]_i$ for $i,\ldots,m^2$ lies within the space spanned by the eigenfunction $\varphi_{j}$, we can expand the vector-valued $\bm{g}_{\bm G}$ in terms of these eigenfunctions as $\bm{g}_{\bm G} (\bm{x})=\sum_{j=1}^{\infty }{\varphi_{j}(\bm{x})\bm{\psi}_{j}}$, where $\bm{\psi}_{j}$ is a set of vector coefficients. Through the iterative applications of $\mathcal{K}_{\bm{G}}$, the following equation is obtained:
\begin{flalign}\label{eq:KSAmatrix}
\bm{g}_{\bm G}(\bm{x}_t)
=(\bm{g}_{\bm G}\circ \underbrace{\bm{f}\circ\cdots\circ\bm{f}}_t)\left(\bm{x}_0\right)
=\sum_{j=1}^{\infty}{\lambda}_j^t{\varphi}_j\left(\bm{x}_0\right)\bm{\psi}_j.
\end{flalign}

For the practical implementation of the spectral decomposition of the linear operator, it is needed to project data onto directions that are effective in capturing the properties of data such as in exact DMD\cite{Tu14}, DMD with reproducing kernels\cite{Kawahara16} and tensor-based DMDs\cite{Klus18,Fujii19b}.
Here, we compute the projection onto some orthogonal directions based on tensor-train decomposition\cite{Oseledets11}, in the same ways of modified tensor-based DMD\cite{Fujii19b}.
In other words, Graph DMD assumes that the data are sufficiently rich and thus a set of the orthogonal direction gives a good approximation of the representation with the eigenfunctions of $\mathcal{K}_{\bm{G}}$. 
Concretely, for an implementation of the above analysis, we first regard the given or calculated matrices as a realisation of the structure of matrix-valued observable $\bm{G}(\bm{x}_t)$ at each $t$. 
We denote the realised matrices (i.e. adjacency matrix) as $\bm{A}_t \in \R^{m\times m}$ for $t=0,\ldots,\tau$. 
Second, we need to compute the projection onto orthogonal directions $\bm{M}$ (see below and Results) and obtain DMD solution $\hat{\bm{F}} \in \R^{p \times p}$.
Then, after eigendecomposition of $\hat{\bm{F}}$, we compute DMD eigenvalues $\hat{\lambda}_j$ and DMD modes $\bm{z}_j \in \C^{m^2}$ for $j = 1,\ldots,p$. 
Finally, we obtain Graph DMD modes $\bm{Z}_j \in \C^{m \times m}$ by matricising $\bm{z}_j$. 

It should be noted that, although we vectorise $\bm{G}(\bm{x}_t)$ in this formulation, the order of the vector is unique; thus it reflects the dependent structure of the matrix-valued observable, $\bm{G}(\bm{x}_t)$. 
In the implementation, we utilise tensor-train decomposition to obtain the matrix $\bm{M} = \bm{X}\bm{N}^\dagger \bm{\Sigma}^{-1} $, where $\bm{X} = [\vectorize(\bm{A}_0),\ldots,\vectorize(\bm{A}_{\tau-1})]$ (see Results).
Although $\bm{A}_t$ is also vectorised, unlike SVD, $\bm{M}$ reflects order-$3$ tensor structure based on tensor-train decomposition.
Therefore, both Koopman spectral analysis with dependent structure among observables and Graph DMD can be calculated without breaking the dependent structure.

\subsection*{Computation of Graph DMD.}
\label{sec:gdmd}

In this subsection, we describe the implementation of Graph DMD\cite{Fujii19b}.
The procedure to compute Graph DMD modes is described in Results. 
The Matlab code we used is available at \url{https://github.com/keisuke198619/GraphDMD}.
To compute this, a modified tensor-based DMD \cite{Fujii19b,Klus18} using tensor-train decomposition\cite{Oseledets11} is applied.
The tensor-train decomposition is considered to be relatively stable and scalable for high-order tensors compared with the other tensor decomposition methods\cite{Oseledets11}. 
The basic algorithm of tensor-train decomposition for an order-$N$ tensor (i.e. $ {\mathcal{A}} \in \mathbb{C}^{n_1 \times \dots \times n_N} $, where $n_l$ denotes the dimensionality of the $l$-th mode for $l = 1,\ldots,N$) decompose into $N$ core tensors $ {\mathcal{A}} ^{(l)} \in  \mathbb{C}^{r_{l-1} \times n_l \times r_l}$, where $r_0 = r_N = 1 $, by serial matricisations and SVDs.
The input tensor of Graph DMD is an order-3 tensor; thus it is decomposed into a matrix, an order-3 tensor, and a matrix.  

For obtaining the time dynamics shown in Eq. \eqref{eq:estimatedGDMD} and Figs. \ref{fig:Diagram} and \ref{fig:example}, we need to compute ${b}_{j,0}$.
Since here we do not need to consider the tensor structure in the time dynamics (i.e. they are univariate time-series), we compute ${b}_{j,0}$ in a similar way of exact DMD by matrisising the DMD mode ${\mathcal{Z}}_{:, :, j} = {\bm{Z}}_j$ (such that the row is the time stamp) and vectorising the initial value $\bm{A}_0$.
After computing the time dynamics, we can compute the reconstructed data shown in Eq. \eqref{eq:estimatedGDMD}.
Using the reconstructed data, we eliminated invalid sliding windows (described below) and investigated the validity of the decomposition.
In this paper, as a criterion of valid decomposition, we used variability accounted for (VAF) which is commonly used in non-orthogonal dimensionality reduction\cite{Fujii19}. 
VAF is defined as the square error of the reconstructed data and the original data such that ${\rm{VAF}}_j = 1-(\|\bm{A} - \hat{\bm{A}}_j \|_F^2)/\|\bm{A}\|^2_F$, where $ \bm A$ and $\hat{\bm{A}}$ are matrisised input data tensor $\mathcal A$ (such that the row is time) and its reconstructed tensor for the $j$the mode, and $\| \, \cdot \, \|_F$ is the Frobenius norm. 
In this paper, if the maximum $\rm{VAF}_j$ < 0.01, the sliding window was eliminated from the subsequent analyses.
We confirmed all attack-segments have at least one valid sliding window.

Next, we investigated the validity of the decomposition when changing the parameters (e.g. the Graph DMD tolerance, sliding window size and cutoff frequency) using a validation dataset.
Here we describe the summary (for the details, see Supplementary Text 2).
First, we set Graph DMD tolerance $\varepsilon$ (i.e. the tolerance in the successive SVD in tensor-train decomposition) to $1.0 \times 10^{-5}$ based on the result in Supplementary Fig. 1a.
For the sliding temporal windows, we set the window size to 50 frames (2 s) including overlaps of 25 frames (1 s) based on the result in Supplementary Fig. 1b.
Regarding the cutoff frequency, we set the cutoff frequency to 2 Hz based on the result in Supplementary Fig. 1c.
Next, we describe the detailed computation of the feature vectors in classification using the Graph DMD modes.

\subsection*{Feature vectors using Graph DMD modes.}
We first adjusted the computed DMD modes.
One is a normalisation within the DMD modes such that the maximum absolute value of the elements is 1 because the amplitude of the modes depends on that of the time dynamics.
Second is symmetrisation. 
Since the computed Graph DMD modes were different from symmetric matrices in a precise sense (although the ideal modes are symmetric matrices), we added the modes to the transposed modes and divided them by two.
Finally, to evaluate the global complex behaviours, we averaged the DMD modes for all valid sliding windows.

\subsection*{Classification.}

After Graph DMD, we perform classification in three different approaches.
For details, see Supplementary Text 1.
The first two approaches create feature vectors and the third automatically extracts the features via a neural network approach.
The final two approaches computed graph features using existing methods: the second is graph Laplacian eigenvalues \cite{Chung97} as a baseline feature of a graph \cite{deLara18} (denoted GDMD Laplacian) and the third is deep graph convolutional neural networks \cite{Zhang18} (denoted GDMD GCN) as a recent promising method for classifying the graph data.
For all classification methods with the exception of the neural network approaches, we adopted logistic regression as a simple linear binary classification model.

As comparable methods for classifying graph sequence data, we adopted four methods.
The first is a method using vectorised basic DMD modes\cite{Tu14} as a baseline of DMD approaches.
The second is the Koopman spectral kernel\cite{Fujii17} using DMD with reproducing kernels\cite{Kawahara16} as the existing method\cite{Fujii18} for classifying the collective motion dynamics. 
The third is a hand-crafted feature as a simple baseline method.
Fourth is an advanced neural network approach for classifying graph sequence data called spatio-temporal GCN \cite{Yan18} as a recent promising end-to-end method.
For the detailed setups, see Supplementary Text 1.

\subsection*{Statistical analysis.}
For an accurate evaluation of classification performance from the multiple viewpoints, in addition to accuracy, AUC and F-measure were calculated to compare the classification performance.
AUC is based on the ROC curve, which is generated by plotting a cumulative distribution function of the true-positive rate with respect to the false-positive rate. 
Then, AUC takes various decision thresholds into consideration.
Next, recall and precision rate were computed for F-measure.
Recall rate is defined as the ratio of the sum of true positives and true negatives to the number of true positives (the true-positive rate), and the precision rate is
defined as the ratio of the sum of true positives and true negatives to false positives.
The trade-off curve between recall and precision was created using the cumulative distribution function.
To evaluate the trade-off, the F-measure was calculated as follows:  ($2 \times$ precision rate $\times$ recall rate) $/$ (precision rate $+$ recall rate).

For investigating the robustness of the classification results, we repeated test sessions for logistic regression five times using different test sets in analogy to five-fold cross-validation (i.e. classified 25 times). 
To compare the various methods, since the hypothesis of homogeneity of variances between methods was rejected with Levene�fs test, the Kruskal-Wallis nonparametric tests were performed. As the post-hoc comparison, Wilcoxon rank sum test with Bonferroni correction was used within the factor where a significant effect in Kruskal-Wallis test was found.
Since comparisons were only performed between GDMD spectrum and others based on our hypothesis, $p$-value was multiplied by six.
We used $r$ values as the effect size for Wilcoxon rank sum test.

We adopted logistic regression for simply investigating the effect of each element of the feature vector. We calculated the odds ratio (related to effect size), its $p$-value and 95\% confidence interval. 
For all statistical calculations, $p< 0.05$ was considered significant.
All statistical analyses were performed using the MATLAB 2018a Statistics and Machine Learning Toolbox (The MathWorks, Inc., MA, USA).


\section*{Acknowledgements (not compulsory)}
This work was supported by JSPS KAKENHI (Grant Numbers 18K18116, 19H04941 and 18H03287) and AMED (Grant Number JP18dm0307009).

\section*{Author contributions statement}
K.F. conceived the study.
K.F. developed and implemented the software with N.T. and Y.K. guidance. 
K.F., M.H. and Y.I. conducted the experiments to obtain the data.
K.F. analysed the results. 
All authors wrote the manuscript. 

\section*{Additional information}
The authors declare that they have no competing interests.

\newpage
\appendix

\section*{Supplementary materials}

\section*{Text 1. Details of classification.}
\subsection*{Classification using Graph DMD modes}

After Graph DMD, we perform classification in three different approaches.
The first two approaches create feature vectors and the last is a neural network approach to automatically create the features.
The first is simply to vectorise the Graph DMD modes (denoted as GDMD spectrum) to directly reflect the node (i.e. players) information. 
We used the elements of the modes regarding the relation among defenders, among attackers and defenders and among defenders and ring (as geometric information) as a feature vector for the team-defense recognition task, and those among attackers and among attackers and defenders (without geometric information) for the team-offense recognition task.
The last two is to compute graph features by existing methods: the second is graph Laplacian eigenvalues \cite{Chung97} as a well-known baseline feature of a graph \cite{deLara18} (denoted as GDMD Laplacian) and the third is deep graph convolutional neural networks \cite{Zhang18} (denoted as GDMD GCN) as a recent promising method to classify the graph data.

For a brief explanation of Graph Laplacian, let $\bm{A} \in \R^{m\times m}$ be an adjacency matrix in an undirected graph, which has elements $A_{i,j}$ for $i,j = 1,\ldots,m$.
Using a degree matrix $\bm{D}$, which is a diagonal matrix having elements $D_{i,i} = \sum_{j=1}^m A_{i,j}$, Graph Laplacian matrix is defined as $\bm{L} = \bm{D} - \bm{A}$.
We computed normalised Graph Laplacian defined as $\mathcal{L} = \bm{D}^{-1/2}\bm{L}\bm{D}^{-1/2}$.
We used the averaged Graph DMD modes as the adjacency matrix $\bm A$.
We then obtained $k$ positive smallest eigenvalues of the normalised graph Laplacians\cite{deLara18}.
We set $k = 10$, which means all eigenvalues except $0$ regarding the averaged Graph DMD modes (the number of the nodes $m = 11$).
Then, we used the eigenvalues in ascending order as an input of the classifier.
This approach extracts the graph topology without the node information.

GCN is a generalised convolutional neural network framework to graphs in the spectral domain. 
The GCN we used\cite{Zhang18} keeps more node information and learn the global node topology by sorting a graph's node in a consistent order called SortPooling layer, so that traditional neural networks can be trained on the graphs.
This GCN is closely related to some type of graph kernels based on structure propagation, especially the Weisfeiler-Lehman subtree kernel\cite{Shervashidze11} and propagation kernel\cite{Neumann16}.
The previous work\cite{Zhang18} showed that the GCN achieved highly competitive classification performance with various graph kernels and neural network methods.
Thus, we used this method as a recent promising method to classify the graph data.
We used the averaged Graph DMD modes as an input adjacency matrix and default parameters in open source code \url{https://github.com/muhanzhang/DGCNN}.

For all classification methods except for the neural network approaches, logistic regression was adopted as a linear binary classification model.
For the neural network approaches\cite{Zhang18,Yan18}, we used the default softmax layer as classifiers.

\subsection*{Classification using other methods}

As comparable methods to classify graph sequence data, we adopted four methods.
The first is a method using vectorised basic DMD\cite{Tu14} modes (denoted as DMD spectrum) as a baseline of DMD approaches (the selection of the elements was the same as GDMD spectrum).
The second is the Koopman spectral kernel\cite{Fujii17} using DMD with reproducing kernels\cite{Kawahara16} as the existing method\cite{Fujii18} to classify the collective motion dynamics (denoted as KDMD spectral kernel). 
In the two methods, input data should be matrix thus we reshaped the input adjacency matrix series to the matrix in which rows and columns were temporal stamps and vectorised adjacency matrix, respectively. 
Koopman spectral kernels generalised a kernel\cite{Vishwanathan07} between dynamical systems to nonlinear dynamical systems.
Among the above kernels, we used Koopman kernel of principal angle between the subspaces of the estimated Koopman mode, showing the best discriminative performance~\cite{Fujii17} using the Koopman modes given by DMD with reproducing kernels. 
Regarding DMD with reproducing kernels\cite{Kawahara16}, we adopted the Gaussian kernel and the kernel width was set as the median of the distances from data. 

The third is a hand-crafted feature as a simple baseline method, consisted of vectorised temporal average, maximum and minimum values of the elements of input adjacency matrix series.
Fourth is an advanced neural network approach to classify graph sequence data called spatio-temporal GCN \cite{Yan18} as a recent promising end-to-end method.
Spatio-temporal GCN is a graph-based neural network for action recognition by modeling dynamic skeletons with the joints as graph nodes and natural connectivities in both human body structures and time as graph edges.
That is, the input of the original work is the joint coordinate vectors on the graph nodes.
Multiple layers of spatio-temporal graph convolution operations will generate higher-level feature maps on the graph.
We used the sequence of adjacency matrices as the input and basically used default parameters in opensource code \url{https://github.com/yysijie/st-gcn}, except for the below parameters.
To adjust the model to the collective motion in this study, since indices of players were based on their distance from the ball and teammates, the neighbour edges were defined as the nearest attackers and defenders, and the ring and players. 
Moreover, for higher-level spatio-temporal graph convolution, we adopted spatial configuration partitioning, which shows the best action recognition performance among various partitioning strategy\cite{Yan18}.
The strategy divides the neighbour set into three subsets: the root node itself, centripetal group and centrifugal group. 
This enables us to semantically higher-level spatio-temporal graph convolution.

\newpage
\section*{Text 2. Selection of Graph DMD parameters.}
Here, we describe the selection of parameters regarding Graph DMD (the Graph DMD tolerance, the temporal window size and the cutoff frequency) and quantitatively validated the applicability of Graph DMD to the sport data in terms of reconstruction error using a validation dataset.

We used the absolute reconstruction error defined as
$(1/m^2\tau)\sum_{t=0}^{\tau-1}\sum_{i=1}^{m}\sum_{j=1}^{m}\|A_{i,j,t} - \hat{A}_{i,j,t} \|$, where $\hat{A}_{i,j,t}$ is a reconstructed element corresponding to $A_{i,j,t}$.
In Graph DMD, Graph DMD tolerance $\varepsilon$ (i.e. the tolerance in the successive SVD in tensor-train decomposition) is critical for the data reconstruction. 
We set $\varepsilon = 1.0 \times 10^{-5}$ because Supplementary Fig. 1a shows that the reconstruction error in this tolerance is lower than other values.

Next, we performed Graph DMD in sliding temporal windows, because complex collective motions often transiently change their rules of motions.
For basic DMD, researchers used sliding windows applied to cortical electroencephalogram data \cite{Brunton16a}.
However, there is a trade-off between the reconstruction error and the meaningful dynamical information in the time interval.
If the window size is too small, the reconstruction error may be small but the extracted information may be useless. 
If too large, the extracted information may reflect meaningful information but the reconstruction error may be too large.
Supplementary Fig. 1b shows that when the window size is 60 frames, the reconstruction error greatly increased.
Thus, we set the window size to 50 frames (2 s) including overlaps of 25 frames (1 s).

In addition, a cutoff frequency is also a important parameter. 
In this study, meaningful motion frequency is considered to be in a low-frequency band (e.g. under 2 Hz) rather than a high-frequency band (e.g. over 2 Hz), when considering the distinction among team-defense or offense motions.
Thus, obtained data was first low-pass filtered at the cutoff frequency.
Moreover, since DMDs are not guaranteed to extract the dynamics within the frequency band, we averaged Graph DMD modes within the temporal frequency band using DMD eigenvalues.
The selection of the cutoff frequency also has a trade-off between the reconstruction error and meaningful dynamical information.
Supplementary Fig. 1c shows that when the cutoff frequency was 3 Hz, the reconstruction error greatly increased.
Thus, we set the cutoff frequency to 2 Hz.


\newpage
\section*{Supplementary Figures}
\renewcommand{\thefigure}{S\arabic{figure}}
\renewcommand{\thetable}{S\arabic{table}}
\setcounter{figure}{0}
\setcounter{table}{0}

\begin{figure}[!h]
\centerline{\includegraphics[width=1 \textwidth]{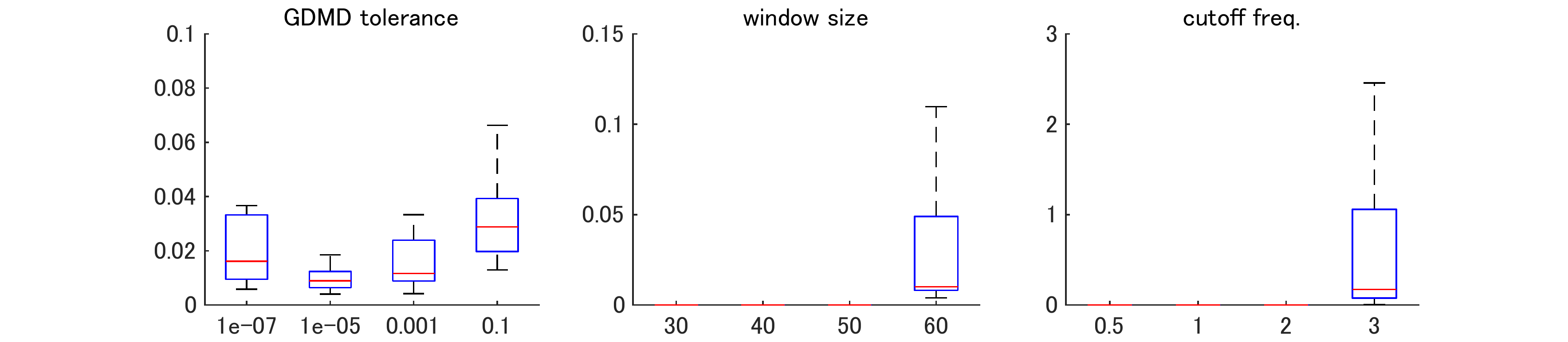}}
\caption{{\bf Reconstruction errors of Graph DMD parameters.}
Reconstruction errors of Graph DMD parameters: (a) Graph DMD tolerance, (b) sliding window size and (c) cut-off frequency of low-pass filter.}
\label{fig:RecError}
\end{figure}
\vspace{20mm}

\begin{figure}[!h]
\centerline{\includegraphics[width=1 \textwidth]{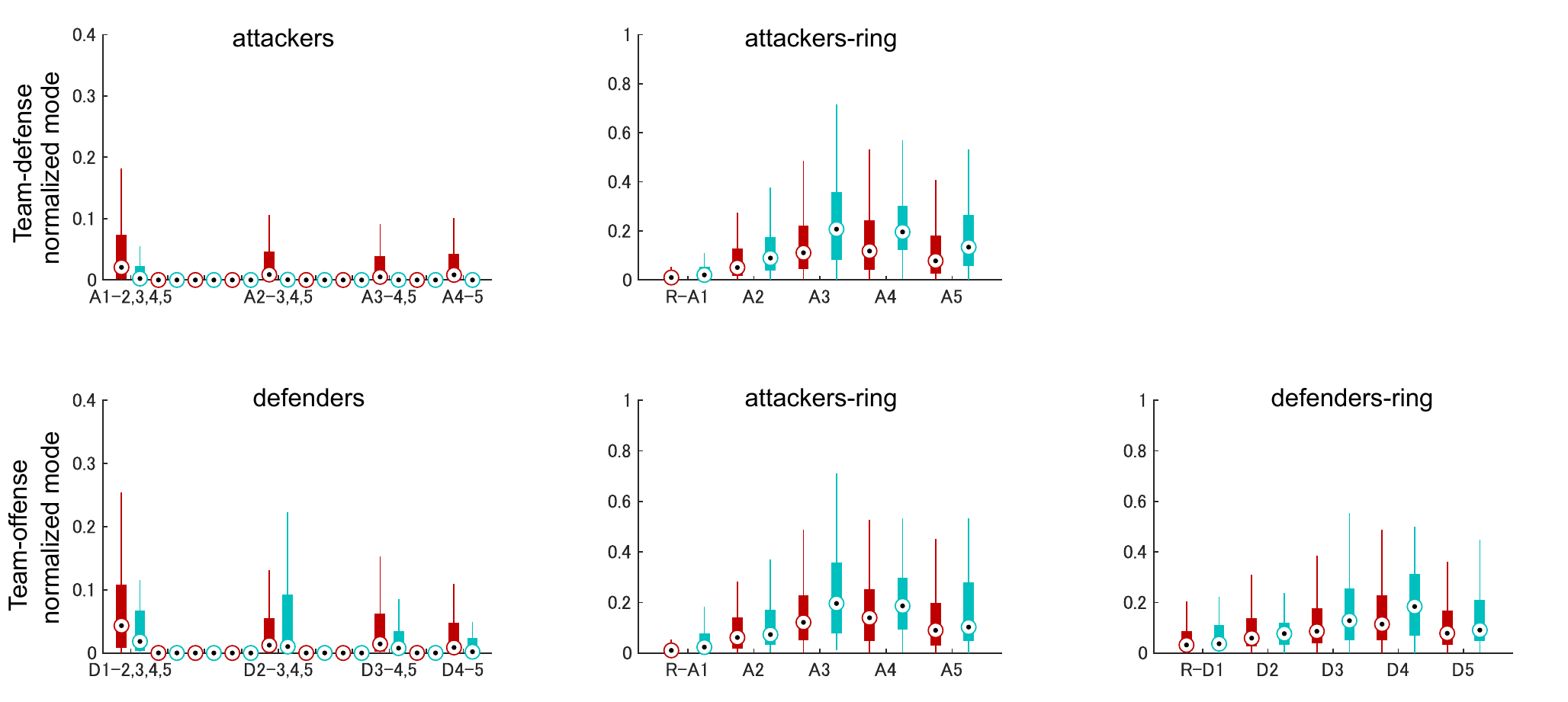}}
\caption{{\bf Boxplots for the elements of Graph DMD modes.}
Boxplots for the elements of Graph DMD modes for label 1 (red) and 2 (blue) which were not used as feature vectors in team-defense (a-c) and offense (d and e) recognition tasks are shown.
These plots are the elements of GDMD modes among attackers (a), and between attackers and the ring (b) in team-defense recognition task, and those among defenders (c), between attackers and the ring (d) and between defenders and the ring (e) in team-offense recognition task.
Notations are same as Fig. 5.}
\label{fig:Boxplot2}
\end{figure}

\newpage
\section*{Supplementary Tables}

\begin{table}[ht!]
\centering
\scalebox{0.6}{
\begin{tabular}{lrrrr}
\hline
& \me{1}{Odds ratio} & \me{1}{$p$} & \me{2}{Confidence interval} \\  
\hline
D1-D2 &$31.815 $ & $0.022$ & $0.492$ & $6.428$ \\D1-D3 & $4.59 \time 10^{9} $ & $0.040$ & $0.990$ & $43.508$ \\D1-D4 & $<0.001 $ & $0.423$ & $-48.727$ & $20.458$ \\D1-D5 & $<0.001 $ & $0.626$ & $-83.331$ & $50.154$ \\D2-D3 & $0.859 $ & $0.927$ & $-3.403$ & $3.100$ \\D2-D4 & $0.038 $ & $0.494$ & $-12.664$ & $6.113$ \\D2-D5 & $<0.001 $ & $0.112$ & $-59.429$ & $6.184$ \\D3-D4 & $6.252 $ & $0.367$ & $-2.153$ & $5.819$ \\D3-D5 & $0.157 $ & $0.633$ & $-9.470$ & $5.763$ \\D4-D5 & $8538.934 $ & $<0.001$ & $4.774$ & $13.331$ \\A1-D1 & $1.485 $ & $0.737$ & $-1.908$ & $2.699$ \\A1-D2 & $0.568 $ & $0.637$ & $-2.915$ & $1.784$ \\A1-D3 & $0.492 $ & $0.532$ & $-2.934$ & $1.516$ \\A1-D4 & $0.029 $ & $0.011$ & $-6.245$ & $-0.805$ \\A1-D5 & $21.537 $ & $0.009$ & $0.752$ & $5.387$ \\A2-D1 & $5.692 $ & $0.175$ & $-0.773$ & $4.251$ \\A2-D2 & $0.785 $ & $0.812$ & $-2.239$ & $1.755$ \\A2-D3 & $1.844 $ & $0.563$ & $-1.462$ & $2.686$ \\A2-D4 & $0.637 $ & $0.686$ & $-2.633$ & $1.732$ \\A2-D5 & $15.120 $ & $0.024$ & $0.362$ & $5.070$ \\A3-D1 & $0.102 $ & $0.045$ & $-4.503$ & $-0.053$ \\A3-D2 & $0.445 $ & $0.505$ & $-3.195$ & $1.574$ \\A3-D3 & $6.438 $ & $0.097$ & $-0.337$ & $4.062$ \\A3-D4 & $0.028 $ & $0.001$ & $-5.719$ & $-1.408$ \\A3-D5 & $1.097 $ & $0.941$ & $-2.362$ & $2.546$ \\A4-D1 & $22.217 $ & $0.017$ & $0.562$ & $5.640$ \\A4-D2 & $0.993 $ & $0.995$ & $-2.185$ & $2.170$ \\A4-D3 & $4.812 $ & $0.187$ & $-0.764$ & $3.907$ \\A4-D4 & $3.795 $ & $0.254$ & $-0.959$ & $3.626$ \\A4-D5 & $2.206 $ & $0.511$ & $-1.566$ & $3.148$ \\A5-D1 & $0.671 $ & $0.736$ & $-2.720$ & $1.922$ \\A5-D2 & $19.102 $ & $0.068$ & $-0.213$ & $6.113$ \\A5-D3 & $1.449 $ & $0.825$ & $-2.911$ & $3.652$ \\A5-D4 & $0.281 $ & $0.398$ & $-4.218$ & $1.676$ \\A5-D5 & $195.490 $ & $<0.001$ & $2.783$ & $7.768$ \\Ring-D1 & $0.002 $ & $<0.001$ & $-9.220$ & $-2.828$ \\Ring-D2 & $0.030 $ & $0.049$ & $-6.966$ & $-0.023$ \\Ring-D3 & $0.136 $ & $0.188$ & $-4.965$ & $0.977$ \\Ring-D4 & $28.668 $ & $0.017$ & $0.593$ & $6.118$ \\Ring-D5 & $0.126 $ & $0.153$ & $-4.912$ & $0.770$ \\
\hline
\end{tabular}}
\caption{\label{tab:Logistic1}{\bf Results of logistic regression in team-defense recognition task.} {\rm }}
\end{table}

\begin{table}[ht!]
\centering
\scalebox{0.6}{
\begin{tabular}{lrrrr}
\hline
& \me{1}{Odds ratio} & \me{1}{$p$} & \me{2}{Confidence interval} \\  
\hline
A1-A2 & $1.06 \time 10^{4} $ & $<0.001$ & $4.770$ & $13.761$ \\A1-A3 & $3.32 \time 10^{30} $ & $0.249$ & $-49.088$ & $189.643$ \\A1-A4 & $7.26 \time 10^{18} $ & $0.635$ & $-136.033$ & $222.890$ \\A1-A5 & $23.986 $ & $0.979$ & $-238.609$ & $244.964$ \\A2-A3 & $11.972 $ & $0.266$ & $-1.894$ & $6.859$ \\A2-A4 & $5.631 $ & $0.792$ & $-11.148$ & $14.605$ \\A2-A5 & $<0.001 $ & $0.219$ & $-481.691$ & $110.421$ \\A3-A4 & $44.656 $ & $0.135$ & $-1.177$ & $8.775$ \\A3-A5 & $0.292 $ & $0.840$ & $-13.176$ & $10.714$ \\A4-A5 & $18.181 $ & $0.135$ & $-0.899$ & $6.700$ \\A1-D1 & $0.139 $ & $0.111$ & $-4.396$ & $0.452$ \\A1-D2 & $0.163 $ & $0.128$ & $-4.152$ & $0.522$ \\A1-D3 & $2.588 $ & $0.391$ & $-1.224$ & $3.125$ \\A1-D4 & $0.080 $ & $0.055$ & $-5.102$ & $0.051$ \\A1-D5 & $23.595 $ & $0.008$ & $0.826$ & $5.496$ \\A2-D1 & $3.565 $ & $0.326$ & $-1.268$ & $3.810$ \\A2-D2 & $0.909 $ & $0.923$ & $-2.009$ & $1.819$ \\A2-D3 & $2.238 $ & $0.459$ & $-1.328$ & $2.940$ \\A2-D4 & $0.668 $ & $0.709$ & $-2.522$ & $1.715$ \\A2-D5 & $1.749 $ & $0.617$ & $-1.630$ & $2.748$ \\A3-D1 & $0.166 $ & $0.102$ & $-3.945$ & $0.359$ \\A3-D2 & $1.341 $ & $0.781$ & $-1.777$ & $2.364$ \\A3-D3 & $3.155 $ & $0.287$ & $-0.966$ & $3.264$ \\A3-D4 & $0.439 $ & $0.447$ & $-2.941$ & $1.296$ \\A3-D5 & $3.807 $ & $0.244$ & $-0.914$ & $3.587$ \\A4-D1 & $5.382 $ & $0.143$ & $-0.571$ & $3.937$ \\A4-D2 & $2.960 $ & $0.334$ & $-1.117$ & $3.288$ \\A4-D3 & $30.457 $ & $0.004$ & $1.112$ & $5.720$ \\A4-D4 & $2.691 $ & $0.397$ & $-1.301$ & $3.281$ \\A4-D5 & $0.994 $ & $0.996$ & $-2.180$ & $2.169$ \\A5-D1 & $0.425 $ & $0.497$ & $-3.327$ & $1.616$ \\A5-D2 & $2.208 $ & $0.603$ & $-2.194$ & $3.778$ \\A5-D3 & $0.036 $ & $0.042$ & $-6.531$ & $-0.116$ \\A5-D4 & $0.695 $ & $0.792$ & $-3.063$ & $2.336$ \\A5-D5 & $26.988 $ & $0.005$ & $0.987$ & $5.604$ \\
\hline
\end{tabular}}
\caption{\label{tab:Logistic2}{\bf Results of logistic regression in team-offense recognition task.} {\rm }}
\end{table}

\end{document}